\documentclass[11pt, letterpaper]{article} 
\usepackage[margin=1in]{geometry}

\usepackage{color}
\usepackage{silence}
\usepackage{cite}
\usepackage{algorithmic}
\usepackage{algorithm}
\usepackage{amssymb}
\usepackage{amsmath}
\usepackage{multirow}
\usepackage{multicol}   
\usepackage[justification=centering]{caption}
\usepackage{footnote}
\makesavenoteenv{tabular}
\usepackage{graphicx}

\usepackage{caption}
\usepackage{subcaption}

\usepackage{epstopdf,float,amsfonts}
\usepackage{epsfig}

\newcommand{\qed}{\hfill \ensuremath{\square}}
\title{\LARGE \bf Reinforcement Learning of Structured Control for Linear Systems with Unknown State Matrix}
\author{Sayak Mukherjee,
           Thanh Long Vu 
\thanks{ S. Mukherjee and T. L. Vu are with the Optimization and Control Group, Pacific Northwest National Laboratory (PNNL), Richland, WA, USA. 
Emails: (sayak.mukherjee, thanhlong.vu)@pnnl.gov.}}
\date{}
\begin{document}

\maketitle
\begin{abstract}
    This paper delves into designing stabilizing feedback control gains for continuous linear systems with unknown state matrix, in which the control is subject to a general structural constraint. We bring forth the ideas from reinforcement learning (RL) in conjunction with sufficient stability and performance guarantees in order to design these structured gains using the trajectory measurements of states and controls. We first formulate a model-based framework using dynamic programming (DP) to embed the structural constraint to the Linear Quadratic Regulator (LQR) gain computation in the continuous-time setting. Subsequently, we transform this LQR formulation into a policy iteration RL algorithm that can alleviate the requirement of known state matrix in conjunction with maintaining the feedback gain structure. Theoretical guarantees are provided for stability and convergence of the structured RL (SRL) algorithm. The introduced RL framework is general and can be applied to any control structure.
    A special control structure enabled by this RL framework is distributed learning control  which is necessary for many large-scale cyber-physical systems. As such, we validate our theoretical results with numerical simulations on a multi-agent networked linear time invariant (LTI) dynamic system.    
\end{abstract}


\section{Introduction}
Researchers over the last decades have been intrigued to find distributed control solutions for interconnected cyber-physical systems, 
in which the controller of each subsystem will receive measurement signals from some neighboring subsystems, instead of the full system state, in order to derive at a decision. On the other hand, in most practical systems, only a subset of the system state is available for the feedback control. There are typical examples of structured feedback control in which a predefined structure is imposed on the feedback control mechanism for practical implementation.
Many research works have been geared toward structured control solutions for different structures and different system models.
Optimal control laws that can capture decentralized \cite{decentralized_review} or a more generic distributed \cite{thesis_distributed} structure has been investigated under notions of quadratic invariance (QI) \cite{QI1, QI2}, structured linear matrix inequalities (LMIs) \cite{dist_LMI1, dist_LMI2}, sparsity promoting optimal control \cite{sparsity_promoting} etc. \cite{geromel} discusses a structural feedback gain computation for discrete-time systems with known dynamics. 

\textcolor{black}{It is worth noting that,} for practical dynamic systems, the dynamic model and its parameters may not always be known accurately (e.g., the US eastern interconnection transmission grid model). Several non-idealities such as unmodeled dynamics from the coupled processes, parameter drift issues over time can make model based control computations insufficient. \textcolor{black}{Unfortunately, most of the aforementioned techniques for the strctured control system design assume that the designer has explicit knowledge of the dynamic system.}
\textcolor{black}{In recent times much attention has been given to model-free decision making of dynamical systems by marrying the ideas from machine learning with the classical control theory, resulting into flourish in the area of reinforcement learning \cite{RL}}. In the RL framework, the control agent tries to optimize its actions based on the interactions with the environment or the dynamic systems quantified in terms of rewards due to such interactions. These techniques were traditionally introduced in the context of sequential decision making using Markov decision processes (MDPs)\cite{Q,bertsekas,ADP1}, and has since been driving force in developing lots of advanced algorithms in RL with applications to games, robotics etc. 

\textcolor{black}{Although many sophisticated machine learning-based control algorithms are developed to achieve certain tasks, these algorithms many times suffer from the lack of stability and optimally guarantees due to multiple heuristics involved in their designs. 
Recent works such as \cite{vrabie1, jiang1} have brought together the good from two worlds to the control of dynamical systems: the capability to learn without model from machine learning and the capability to make decision with rigorous stability guarantee from automatic control theory. Basically, these works leveraged the conceptual equivalence of RL/ADP algorithms with the classical optimal \cite{optimal_book} and adaptive control\cite{adaptive} of the dynamic systems. References \cite{vrabie1, jiang1, jiang_book, V17, V18, sayak_cdc, sayak_arxiv} cover many of such work for systems with partially or completely model free designs.} 

\textcolor{black}{However, the area of  structure based RL/ADP design for dynamic systems is still unexplored to some extent. In \cite{sayak_arxiv, sayak_acc}, a projection based reduced-dimensional RL variant have been proposed for singularly perturbed systems along with a block-decentralized design for two time-scale networks, \cite{Jiang_PS} presents a decentralized design for a class of interconnected systems, \cite{eth_distributed_learning} presents a structured design for discrete-time time-varying systems in the recent times.  Along this line of research, this paper will present a structured optimal feedback control methodology using reinforcement learning without knowing the state dynamic model of system.}  

\par
 We first formulate a model-based constrained optimal control  criterion using the methodologies and guarantees from dynamic programming. Subsequently, we formulate a model-free RL gain computation algorithm that can encode the general structural constraint on the optimal control. This structured learning algorithm - SRL encapsulates the stability and convergence guarantees along with the sub-optimality for the controllers with the specified structure. We substantiate our design on a 6-agent network system.
The paper is organized as follows. We discuss the problem formulation and the required assumptions in the section II. In section III, we discuss the structured RL algorithm development incorporating the structural constraint. Numerical simulation example is given in Section IV, and concluding remarks are given in Section V.
\section{Model and Problem Formulation}
We consider a linear time-invariant (LTI) continuous-time dynamic system:
\begin{align}\label{system}
    \dot{x} = Ax + Bu,\; x(0)=x_0,
\end{align}
where $x \in \mathbb{R}^{n}, u \in \mathbb{R}^{m}$ are the states and control inputs. We, hereby, make the following assumption.\\
\textbf{Assumption 1:} The dynamic state matrix $A$ is unknown.
\par
\begin{figure}[]
    \centering
    \includegraphics[width = .7\linewidth, trim=4 4 4 10, clip]{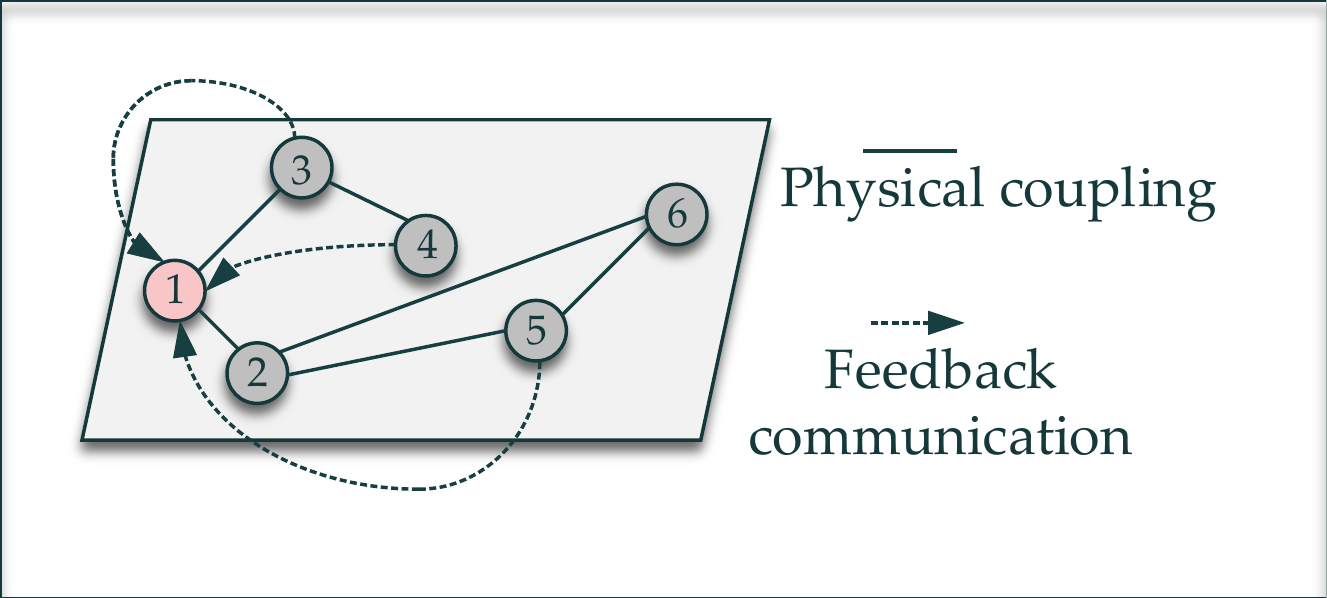}
    \caption{An example of structured feedback for agent $1$}
    \label{fig:scheme}
    \vspace{-.5 cm}
\end{figure}

With this unknown state matrix, we would like to learn an optimal feedback gain $u=-Kx$. However, instead of unrestricted control gain $K \in \mathbb{R}^{m \times n}$, we impose some structure on the gain. We would like to have $K \in \mathcal{K}$, which we call \textbf{\textit{structural constraint}}, where $\mathcal{K}$ is the set of all structured controllers such that: 
\begin{align}\label{struc}
    \mathcal{K} := \{ K \in \mathbb{R}^{m \times n} \; | \; F(K) = 0\}.
\end{align}
Here $F(.)$ is the matricial function that can capture any structure in the control gain matrix. This can encode which elements in the $K$ matrix will be non-zero, for example, with the multi-agent example as given in Fig. \ref{fig:scheme}, the control for the agent $1$ can be constrained in the form:
\begin{align}\label{example_struc}
    K_1 \in \begin{bmatrix} 0 & 0 & * & * & * & 0 \end{bmatrix}.
\end{align}
$K_1$ denotes the first row of $K$. Similarly, the feedback communication requirement on all other agents can be encoded. Here all such feedback gains with the specified structure will be captured in the set $\mathcal{K}$. Therefore, we make the following assumption on the control gain structure.\\
\textbf{Assumption 2:} The communication structure required to implement the feedback control is known to the designer, and it is sparse in nature.
\par
This assumption means that the structure $\mathcal{K}$ is known. This captures the limitations in the feedback communication infrastructure, and can also some time represent a much less dense infrastructure to minimize the cost of deployment of the network. For many network physical systems, the communication infrastructure can be already existing, for example, in some peer-to-peer architecture, agents can only receive feedback from their neighbors. Another very commonly designed control structure is of block-decentralized nature where local states are used to perform feedback control. Therefore, our general constraint set will encompass all such scenarios. We also make the standard stabilizability assumption.\\
\textbf{Assumption 3:} The pair $(A,B)$ is stabilizable and $(A,Q^{1/2})$ is observable.
\par
We can now state the model-free  learning control problem  as follows.\\
\textbf{P.} \textit{For the system \eqref{system} satisfying Assumptions $1, 2, 3$, \textbf{learn}  structured control policies $u = -Kx$, where $K \in \mathcal{K}$ described in \eqref{struc}, such that the closed-loop system is stable and the following objective is minimized}
\begin{align}\label{Ji}
    J (x(0),u^\mathcal{K}) = \int_{0}^{\infty} (x^TQx + u^TRu) d\tau.
\end{align}
    \par

\section{Structured Reinforcement Learning (SRL)}
To develop the learning control design we will take the following route. We will first formulate a model-based solution of the optimal control problem via a modified Algebraic Riccati Equation (ARE) in continuous-time that can solve the structured optimal control when all the state matrices are known. Thereafter, we will formulate a learning algorithm that does not require the knowledge about the state matrix $A$ by embedding the model-based framework with guarantees into RL, which will then automatically ensure the convergence and stability of the learning solution. 
First of all, we present the following central result of this paper.\\
\textbf{Theorem 1:}\textit{ For any matrix $L \in \mathbb{R}^{m \times n}$, 
    let $P \succ 0$ be the solution of the following modified Riccati equation
    \begin{align}\label{struc_ARE}
       A^TP + PA -PBR^{-1}B^TP + Q + L^TRL = 0. 
    \end{align}
    Then, the control gain 
\begin{align}
\label{Lform}
    K= K(L)  = R^{-1}B^TP - L,
\end{align} will ensure closed-loop stability, i.e., $A-BK \in \mathbb{RH}_{\infty}$.
 \qed\\}
\textbf{Proof:}
We look into the optimal control solution of the dynamic system (1) with the objective (2) using dynamic programming (DP) to ensure theoretical guarantees. We assume at time $t$, the state is at $x = x_1$. We define the finite time optimal value function with the unconstrained control as:
\begin{align}
    V_t(x_1) = \mbox{min}_{u} \int_{t}^{T} (x^TQx + u^TRu) d\tau,
\end{align}
with $x(t) = x_1, \dot{x} = Ax + Bu$. Staring from state $x_1,$ the optimal $V_t(x_1)$ gives the minimum LQR cost-to-go. Now as the value function is quadratic, we can write it in a quadratic form as, $V_t(x_1)  = x_1^TP_tx_1, \; P_t \succ 0.$ For a small time interval $[t, t+h]$, where $h >0$ is small, we assume that the control is constant at $u = u_1$ and is optimal. Then cost incurred over the interval $[t, t+h]$ is 
\begin{align}
    U_1 = \int_t^{t+h} (x^TQx + u^TRu) d\tau \approx h(x_1^TQx_1 + u_1^TRu_1).
\end{align}
Also, the control input evolves the states at time $t+h$ to,
\begin{align}
    x(t+h) = x_1 + h(Ax_1 + Bu_1).
\end{align}
Then, the minimum cost-to-go from $x(t+h)$ is:
\begin{align}
    V_{t+h}(x(t+h)) &= (x_1 + h(Ax_1 + Bu_1))^T P_{t+h} (x_1 \nonumber \\  &\;\;\;\;\; +h(Ax_1 + Bu_1)).
\end{align}
Expanding $P_{t+h}$ as $(P_t + h\dot{P}_t)$ we have
   \begin{align} 
   &V_{t+h}(x(t+h)) = \nonumber \\
    &(x_1 + h(Ax_1 + Bu_1))^T (P_t + h\dot{P}_t) (x_1 + h(Ax_1 + Bu_1)),\\
    &\approx x_1^TP_tx_1 + h((Ax_1 + Bu_1)^TP_tx_1 + x_1^TP_t(Ax_1 + Bu_1) \nonumber \\
    &+ x_1^T\dot{P}_tx_1).
\end{align}
Here, we neglect higher-order terms. Therefore, the total cost-
\begin{align}
    V_t(x_1) &= U = U_1 + V_{t+h}(x(t+h)),\\
    &=x_1^TP_tx_1 + h(x_1^TQx_1 + u_1^TRu_1 + (Ax_1 + Bu_1)^TP_tx_1 \nonumber \\ 
    & + x_1^TP_t(Ax_1 + Bu_1) + x_1^T\dot{P}_tx_1).
\end{align}
If the control $u=u_1$ is optimal then the total cost must be minimized. Minimizing \textcolor{black}{$V_t(x_1)$} over $u_1$ we have,
\begin{align}
    u_1^TR + x_1^TP_tB = 0,\\
    u_1 = -R^{-1}B^TP_tx_1.
\end{align}
Now, this gives us an optimal gain $K = R^{-1}B^TP_t$ which solves the unconstrained LQR. However, we are not interested in the unconstrained optimal gain, as that cannot impose any structure as such. In order to impose structure in the feedback gains, the feedback control will have to deviate from the optimal solution of $R^{-1}B^TP_t$, and following \cite{geromel}, we introduce another matrix $L \in \mathbb{R}^{m \times n}$ such that,
\begin{align}
    K+ L =  R^{-1}B^TP_t,\\
    K= R^{-1}B^TP_t - L.
\end{align}
The matrix $L$ will help us to impose the structure, i.e., $K \in \mathcal{K}$, which we will discuss later. Therefore, the structured implemented control is given by,
\begin{align}\label{control1}
    u_1 = -Kx_1 = -R^{-1}B^TP_tx_1 + Lx_1. 
\end{align}
We have $u_1 \in u^{\mathcal{K}}$, where $u^\mathcal{K}$ is the set of all control inputs when following $K \in \mathcal{K}$. Now, with slight abuse of notation, we denote the matrix $P_t$ to be the solution corresponding to the structured optimal control. The Hamilton-Jacobi equation with the structured control is given by,
\begin{align}
    V_t^{K \in \mathcal{K}}(x_1) &\approx \mbox{min}_{u_1 \in u^{\mathcal{K}}} U,\\
    x_1^TP_tx_1 & \approx \mbox{min}_{u_1 \in u^{\mathcal{K}}} (x_1^TP_tx_1 + h(x_1^TQx_1 + u_1^TRu_1 \nonumber \\ 
    &+ (Ax_1 + Bu_1)^TP_tx_1 + x_1^TP_t(Ax_1 + Bu_1) + \nonumber \\
    &x_1^T\dot{P}_tx_1)).
\end{align}
Putting \eqref{control1}, neglecting higher order terms, and after simplifying we get,
\begin{align}
    -\dot{P}_t = A^TP_t + P_tA -P_tBR^{-1}B^TP_t + Q + L^TRL.
\end{align}
For steady-state solution, we will have,
\begin{align}
    A^TP + PA -PBR^{-1}B^TP + Q + L^TRL  = 0. 
\end{align}
This proves the modified Riccati equation of the theorem. Now let us look into the stability of the closed-loop system with the gain $K = R^{-1}B^TP - L$. We can consider the Lyapunov function:
\begin{align}
    W = x^TPx, \; P \succ 0.
\end{align}
Therefore, the time derivative along the closed-loop trajectory is given as,
\begin{align}
    \dot{W} &= x^TP\dot{x} + \dot{x}^TPx,\\
    &= x^TP(Ax + B(-R^{-1}B^TPx + Lx)) + \nonumber. \\
    &\;\;\; (Ax + B(-R^{-1}B^TPx + Lx))^TPx,\\
    &= x^T[PA + A^TP - PBR^{-1}B^TP - PBR^{-1}B^TP \nonumber \\ &\;\;\;\; +PBL + L^TB^TP]x,\\
    &= x^T[-Q  - PBR^{-1}B^TP + 2PBL - L^TRL]x,\\
    &= x^T[-Q - (PBR^{-1}-L^T)R(PBR^{-1}-L^T)^T]x.
\end{align}
Now as $R$ is positive definite, the terms of form $X^TRX$ are at-least positive semi-definite. Therefore, we have,
\begin{align}
    \dot{W} \leq -x^TQx.
\end{align}
This ensures, closed-loop system stability. For the linear system, with the assumption that $(A,Q^{1/2})$ is observable, the globally asymptotic stability can be proved \textcolor{black}{by using the LaSalle's invariance principle}. This completes the proof. \qed
\par
 At this point, we investigate closely the matricial structure constraint. Let $I_{\mathcal{K}}$ denotes the indicator matrix for the structured matrix set $\mathcal{K}$ where this matrix contains element $1$ whenever the corresponding element in $\mathcal{K}$ is non-zero. For the example \eqref{example_struc}, we will have for the first row as,
     \begin{align}\label{example_indicator}
    I_{\mathcal{K}_1} = \begin{bmatrix} 0 & 0 & 1 & 1 & 1& 0\end{bmatrix}.
\end{align}
 
 Therefore structural constraint is simply written as:
\begin{align}\label{struc2}
    F(K) = K \circ I^c_{\mathcal{K}}=\bf{0}.
\end{align}
Here, $\circ$ denotes the element-wise/ Hadamard product, and $I^c_{\mathcal{K}}$ is the complement of $I_{\mathcal{K}}$. We, hereafter, state the following theorem on the choice of $L$ to impose structure on $K$. This follows the similar form of discrete-time condition of \cite{geromel}. \\


\textbf{Theorem 2:} Let $P$ be the solution of the modified ARE \eqref{struc_ARE} where
$
    L = F(\phi(P))$ and $\phi(P) = R^{-1}B^TP.
$ Then, the control gain $K=R^{-1}B^TP - L$ designed as in Theorem 1 
will satisfy the structural constraint $F(K) = 0.$

\noindent \textbf{Proof}: We have,
\begin{align}
    K &= \phi(P) - L = \phi(P) - F(\phi(P))\\
    &= \phi(P) - \phi(P)\circ I^c_{\mathcal{K}}\\
    &= \phi(P) \circ (\mathbf{1}_{m \times n} - I^c_{\mathcal{K}} )\\
    &= \phi(P) \circ I_{\mathcal{K}} \in \mathcal{K}.
\end{align}
This concludes the proof. \qed\\ 
    
We note that the implicit assumption here is the existence of the solution of the modified ARE \eqref{struc_ARE} where
$L = F(\phi(P))$ and $\phi(P) = R^{-1}B^TP.$ It is still an open question on the necessary and sufficient condition on the structure $F(K)$ for the existence of this solution. However, once the solution exists, we can 
 iteratively compute it and the associated control gain $K$ using the following model-based  algorithm. \par  
\noindent \textbf{Theorem 3: } \textit{Let $K_0$ be such that $A-BK_0$ is Hurwitz. Then, for $ k=0,1,\dots $ \\
1. Solve for ${P}_k$(Policy Evaluation) :
\begin{align}\label{Kleinman1}
\hspace{-.3 cm} &A_{k}^T{P}_k + {P}_k A_{k} + {K}_k^TR{K}_k + Q = 0, A_{k} = A-B{K}_k.
\end{align}
2. Update the control gain (Policy update):
\begin{align}\label{Kleinman2}
{K}_{k+1} = R^{-1}B^T {P}_k - F(\phi(P_k)), \; \phi(P_k) = R^{-1}B^T P_k.
\end{align}
Then $A - BK$ is Hurwitz and $K_{k} \in \mathcal{K}$ and $P_{k}$ converge to structured $K \in \mathcal{K}$, and $P$ as $ k  \rightarrow \infty $.} 
\qed \par
\noindent \textbf{Proof:} The theorem is formulated by taking motivation from the Kleinman's algorithm \cite{kleinman} that has been used for unstructured feedback gain computations. Comparing with the Kleinman's algorithm, the control gain update step is now modified to impose the structure. With some mathematical manipulations it can be shown that using $K_k= R^{-1}B^TP_k -F_k$, where $
    F_k = R^{-1}B^TP_k \circ I_{\mathcal{K}}^c,
$ we have,
\begin{align}
   &(A-BK_k)^TP_k + P_k(A-BK_k) + K_k^TRK_k = \\
    & A^TP_k + P_kA -P_kBR^{-1}B^TP_k + F_k^TRF_k = -Q. 
\end{align}
Therefore, \eqref{Kleinman1} is equivalent to \eqref{struc_ARE} for the $k^{th}$ iteration. As we shown the stability and convergence via dynamic programming and Lyapunov analysis in Theorem $1$, considering the equivalence of Theorem $1$ with this iterative version, the theorem can be proved.   \qed
\par
\par
Although, we have formulated an iterative solution to compute structured feedback gains, the algorithm is still model-dependent. We, hereby, start to move into a state matrix agnostic design using reinforcement learning. 
We write \eqref{system} incorporating $u = -K_kx, K_k \in \mathcal{K}$ as
\begin{align}
\dot{x} &= Ax + Bu = (A - BK_k)x + B(K_kx + u) .
\end{align} 
We \textit{explore} the system by injecting a probing signal $u = u_0$ such that the system states do not become unbounded  \cite{jiang_book}. For example, following \cite{jiang_book} one can design $u_0$ as a sum of sinusoids. Thereafter, we consider a quadratic Lyapunov function $x^TPx, P \succ 0$, and we can take the time-derivative along the state trajectories, and use Theorem $3$ to alleviate the dependency on the state matrix $A$. 
\begin{align}
&\frac{d}{dt}(x^TP_kx) = x^T(A_{k}^TP_k + P_kA_{k})x + \nonumber \\ 
& \;\;\;\;\;\;\;\;\;\;\;\;\;\;\;\;\;\;\;\;\;\;\; 2(K_k x+u_{0})^TB^TP_k x, \\ 
& = -x^T(\bar{Q}_{k} +  2(K_kx+u_0)^TR(K_{i(k+1)}+F_k))x, \nonumber 
\end{align}
where, $\bar{Q}_{k} = Q + K_k^TRK_k $. Therefore, starting with an arbitrary control policy $u_0$ we have,
\footnotesize
\begin{align}
\label{main eqn}
 &\hspace{-.28 cm} x^T_{(t+T)}P_kx_{(t+T)} - x^T_t P_k x_t  
 -  2\int_{t}^{t+T}((K_kx+u_{0})^TR (K_{k+1}+F_k)x )d\tau  \nonumber \\ & \;\;\;\;\; = -\int_{t}^{t+T}(x^T \bar{Q}_k x)d\tau.
 \end{align}
 \normalsize
 We, thereafter, solve \eqref{main eqn} by formulating an iterative algorithm using the measurements of the state and control trajectories. The design will require to gather data matrices $\mathcal{D} = \{ S_{xx}, T_{xx}, T_{xu_0}\}$ for sufficient number of time samples (discussed shortly) where,
 \begin{align} 
& \hspace{-.3 cm} S_{xx} = \begin{bmatrix}
x \otimes x |_{t_1}^{t_1+T},& \cdots &, x \otimes x |_{t_l}^{t_l+T} 
\end{bmatrix}^T,\\
& \hspace{-.3 cm} T_{xx} = \begin{bmatrix}
\int_{t_1}^{t_1+T}(x \otimes x) d\tau ,& \cdots &, \int_{t_l}^{t_l+T} (x \otimes x) d\tau \\
\end{bmatrix} ^T,\\
& \hspace{-.3 cm} T_{xu_0} = \begin{bmatrix}
\int_{t_1}^{t_1+T}(x \otimes u_0) d\tau ,& \cdots & ,\int_{t_l}^{t_l+T} (x \otimes u_0) d\tau \\
\end{bmatrix} ^T.
\end{align} 
\vspace{-.4 cm}
 \begin{algorithm}[H]
\footnotesize
\caption{ Structured Reinforcement Learning (SRL) Control}
1. \textit{Gather sufficient data:}
\textit{Store} data ($x,$ and $u_0$) for interval $(t_1,t_2,\cdots,t_l),t_i-t_{i-1}=T$.
Then \textit{construct} the following data matrices $\mathcal{D} = \{ S_{xx}, T_{xx}, T_{xu_0}\}$ such that rank($T_{xx} \;\; T_{xu_0}) = n(n+1)/2 + |\mathcal{K}|$. \\

2. \textit{Controller update iteration :}
Starting with a stabilizing $K_0$, \textit{Compute} $K$ iteratively ($k=0,1,\cdots$) using the following iterative equation

\textbf{for $k=0,1,2,..$}\\
A. \textit{Solve} for $P_k,$ and  $K_{k+1}+F_k$:
\begin{align}\label{eq:update}
 \underbrace{\begin{bmatrix}
S_{xx} & -2T_{xx}(I_n \otimes K_k^TR)  -2T_{xu_0}(I_n \otimes R)
\end{bmatrix}}_{\Theta_k}\begin{bmatrix}
vec(P_k) \\ vec(K_{k+1} + F_k) 
\end{bmatrix}   =\underbrace{-T_{xx}vec(\bar{Q}_{k})}_{\Phi_k}.
\end{align}
B. \textit{Compute} $F_k = R^{-1}B^TP_k \circ I_{\mathcal{K}}^c $ using the feedback structure matrix.\\ 
C. \textit{Update} the gain $K_{k+1}$.\\
D. \textit{Terminate} the loop when $||P_k - P_{k-1}|| < \varsigma$, $\varsigma > 0$ is a small threshold.\\
\textbf{endfor}\\

3. \textit{Applying K on the system :} Finally, apply $u=-Kx, K \in \mathcal{K}$, and remove $u_0$.\\
\end{algorithm} 
\normalsize

\normalsize
 Algorithm $1$ presents the steps to compute the structured feedback gain $K \in \mathcal{K}$ without knowing the state matrix $A$.

\noindent \textbf{Remark 1:} If $A$ is Hurwitz, then the controller update iteration in \eqref{eq:update} can be started without any stabilizing initial control. Otherwise, stabilizing $K_0$ is required, as commonly encountered in the RL literature \cite{jiang_book}. This is mainly due to its equivalence with modified Kleinman's algorithm in Theorem $3$.\par 
\noindent \textbf{Remark 2:} The rank condition dictates the amount of data sample needs to be gathered. For this algorithm we need  rank($T_{xx} \;\; T_{xu_{0}}) = n(n+1)/2 + |\mathcal{K}|$, where $|\mathcal{K}|$ is the number of non-zero elements in the structured feedback control matrices. This is based on the number of unknown variables in the least squares. The number of data samples can be considered to be twice this number to guarantee convergence.
\par
The next theorem describes the sub-optimality of the structured solution with respect to the unconstrained objective.\\
\textbf{Theorem 4:} The difference between the optimal structured control objective value $J$ and  the optimal  unstructured objective $\bar{J}$ is bounded as:
\begin{align}
\label{eq.bound}
    \| J - \bar{J}\|  \le \frac{l}{2g} \| (x_0^T \otimes x_0^T)\|,
\end{align}
for  any  control structure. Here $g=\|BR^{-1}B^T\|_2,$ $l = \| \mathbf{V}^{-1}\|^{-1}$, where $\mathbf{V}: \mathbb{R}^{n} \to \mathbb{R}^n$ is a operator defined as:
\begin{align}
    \mathbf{V}W = (A - BR^{-1}B^T)^TW + W(A - BR^{-1}B^T).
\end{align} \qed \\
\noindent \textbf{Proof:} Let the unstructured solution of the ARE be denoted as $\bar{P}$, then the unstructured objective value is $\bar{J} = x_0^T\bar{P}x_0$, whereas, the learned structured control will result into the objective $J = x_0^TPx_0$, therefore we have,
\begin{align}
    \| J - \bar{J}\| &= \| (x_0^T \otimes x_0^T) vec (P - \bar{P})\|  \nonumber \\
    & \leq \| (x_0^T \otimes x_0^T)\| \|(P - \bar{P})\|_F 
\end{align}
Using $g$, and $l$ as defined in the Theorem 4 statement, following \cite[Theorem 3]{are_pert} 
with $\epsilon = \frac{\|L^TRL\|}{l}$, we have,
\begin{align*}
    \|(P - \bar{P})\|_F &\leq \frac{2l\epsilon}{l + \sqrt{l^2 - 4lg\epsilon}} = \frac{2l\epsilon(l - \sqrt{l^2 - 4lg\epsilon})}{ 4lg\epsilon} \\
          & = \frac{(l - \sqrt{l^2 - 4lg\epsilon})}{ 2g}
          <\frac{l}{2g}.
\end{align*}

As such, the difference between the optimal values $J$ and $\bar{J}$ is bounded by
\begin{align}
\label{eq.bound}
    \| J - \bar{J}\|  \le \frac{l}{2g} \| (x_0^T \otimes x_0^T)\|
\end{align}
for any structure of the control. We note that $g$ and $l$ are not dependent on the control structure. Therefore, the inequality \eqref{eq.bound} indicates that the difference between the optimal control value $J$ with $K \in \mathcal{K}$, and optimal unstructured control value $\bar{J}$ is linearly bounded by the Kronecker combination of the initial value  $\| (x_0^T \otimes x_0^T)\|$ for any control structure.\qed

\section{Numerical Example}
We consider a multi-agent network with $6$ agents following the interaction structure shown in Fig. \ref{fig:scheme}. We consider each agent to follow a consensus dynamics with its neighbors such that:
\begin{align}
    \dot{x}_i = \sum_{j \in \mathcal{N}_i, i \neq j} \alpha_{ij}(x_j - x_i) + u_i, x_i(0) = x_{i0}, 
\end{align}
where $\alpha_{ij} > 0$ are the coupling coefficients. We consider the state and input matrix to be:
\begin{align}
\footnotesize
    A = \begin{bmatrix} -5 &2& 3& 0& 0& 0\\
     2 &-6 &0& 0& 1& 3\\
     3 &0& -5& 2 &0 &0\\
     0 &0& 2& -2& 0& 0\\
     0 &1& 0 &0 &-4 &3\\
     0 &3& 0 &0& 3& -6 \end{bmatrix}, B=I_6.
\end{align}
\normalsize
The dynamics given as above is generally referred to as a Laplacian dynamics with $A.\mathbf{1_n} = \mathbf{0}$ resulting into a zero eigenvalue. We would like the controller to improve the damping of the eigenvalues closer to instability. The eigenvalues of the system are $-10.00,
   -8.27,
   -6.00,
   -3.00,
   -0.72,
   -0.00$. We choose initial conditions as $[0.3,0.5,0.4,0.8,0.9,0.6]^T$. We consider two scenarios with structured gains: A. $I_\mathcal{K}(1,2) = 0, I_\mathcal{K}(1,6) = 0, I_\mathcal{K}(1,1) = 0, I_\mathcal{K}(2,4) = 0,I_\mathcal{K}(2,6) = 0,
I_\mathcal{K}(3,4) = 0, I_\mathcal{K}(3,5) = 0$, and B. Along with the sparsity pattern in scenario A, we also have $ I_\mathcal{K}(4,1) = 0, I_\mathcal{K}(4,2) = 0, I_\mathcal{K}(5,4) = 0, I_\mathcal{K}(5,3) = 0,
I_\mathcal{K}(6,4) = 0,
I_\mathcal{K}(6,1) = 0$. Please note that we consider an arbitrary sparsity pattern. We assume that the states of all the agents can be measured. 

We consider the design parameters as $Q = 30I_6, R = I_6$. First we describe the scenario $A$. Here we have $n=6, m=6$, and number of non-zero elements of $K$ is $29$, therefore, we require to gather data for at-least $2(n(n+1)/2 + |\mathcal{K}|)$ samples, which is $100$ data samples. We consider the time step to be $0.01 $s, and gather data for $1.4$ s. The iteration for $K$ and $P$ took around $0.02$ s on an average with a Macbook laptop of Catalina OS, 2.8 GHz Quad-Core Intel Core i7 with 16 GB RAM. During exploration, we have used sum of sinusoids based perturbation signal to persistently excite the network. Please note that the majority of the learning is spent on the exploration of the system because of the requirement of persistent excitation and  the least square iteration is a order of magnitude smaller in comparison to the exploration time. With faster processing units, the least square iteration can be made much faster. Fig. \ref{fig:c1_traj} shows the state trajectories of the agents during exploration, and also with control implementation phase. The structured control gain learned in this scenario is given as:

\footnotesize
\begin{align}
    K^A_\mathcal{K}= \begin{bmatrix}
    0.0000  & 0.0000 &   1.2527   & 0.2901   & 0.1468  &  0.0000\\
    1.0455 &   2.7516  &  0.2686  &  0.0000  &  0.7485  &  0.0000\\
    1.2527  &  0.2686  &  2.9976  &  0.0000 &  0.0000  &  0.0670\\
    0.2901  &  0.0364  &  1.0471 &   4.1729 &   0.0025 &   0.0054\\
    0.1468  &  0.7485  &  0.0288 &   0.0025 &   3.2813 &   1.2978\\
    0.3306  &  1.1411  &  0.0670  &  0.0054 &   1.2978  &  2.8851
    \end{bmatrix}
\end{align}
\normalsize
The total cost comes out to be $12.4705$ units. Fig. \ref{fig:c1p}-\ref{fig:c1k} show that the $P$ and $K$ iteration converges after around $6$ iterations. The structured solution also matches with the model-based solution from Theorem 3 with high accuracy. Whenever, the learning has been performed for unstructured LQR control gain, the solution comes out to be:
\begin{figure}[t]
    \centering
    \begin{minipage}{\linewidth}
        \centering
        \includegraphics[width = .75\linewidth, height = 4.5 cm]{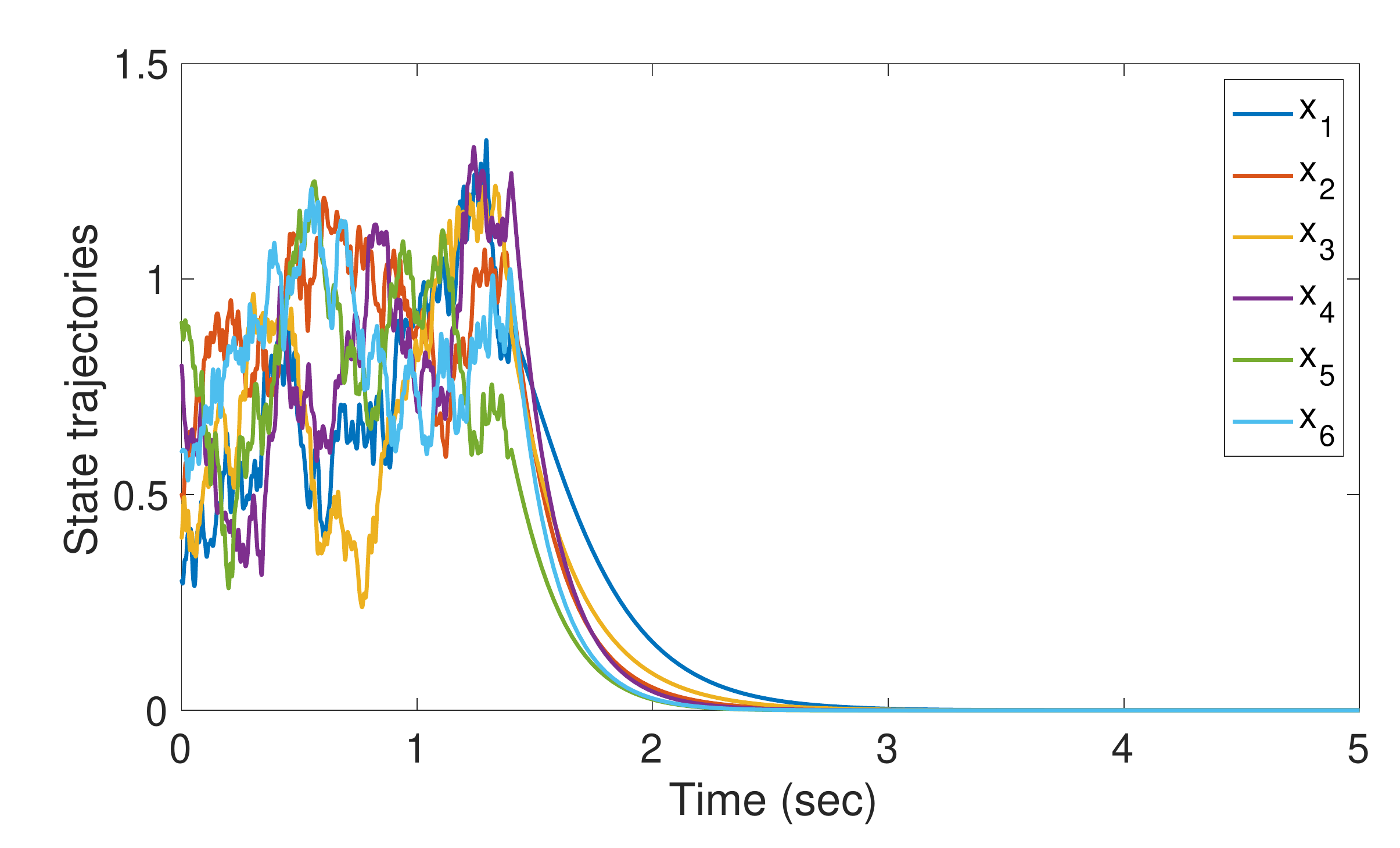} 
        \caption{Scenario A: State trajectories during exploration and control implementation}
        \label{fig:c1_traj}
    \end{minipage}
    \centering
    \begin{minipage}{0.4\linewidth}
        \centering
        \includegraphics[width = \linewidth]{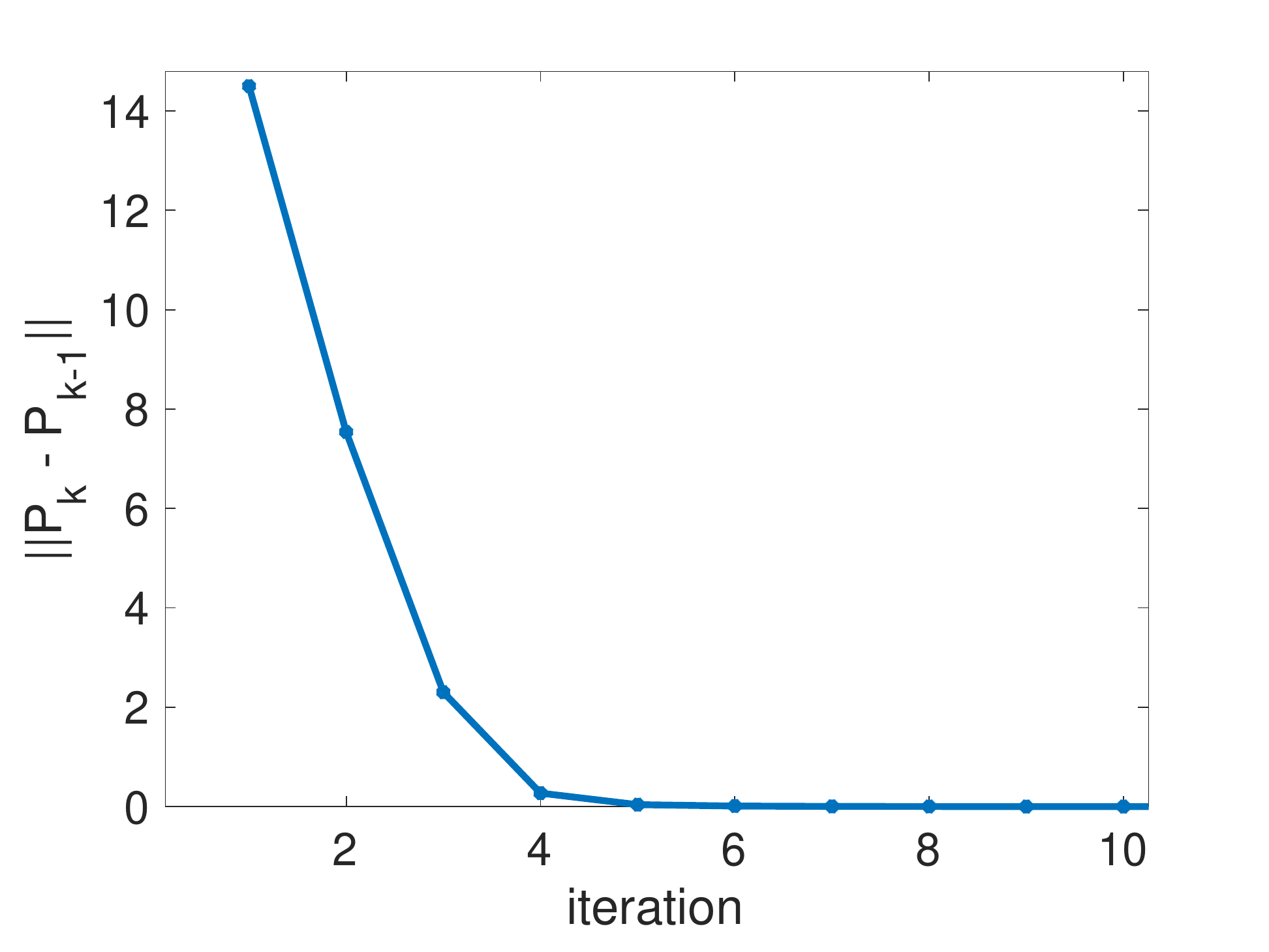} 
        \caption{Scenario A: P convergence}
        \label{fig:c1p}
    \end{minipage}
    \quad
    \begin{minipage}{0.4\linewidth}
        \centering
       \includegraphics[width = \linewidth]{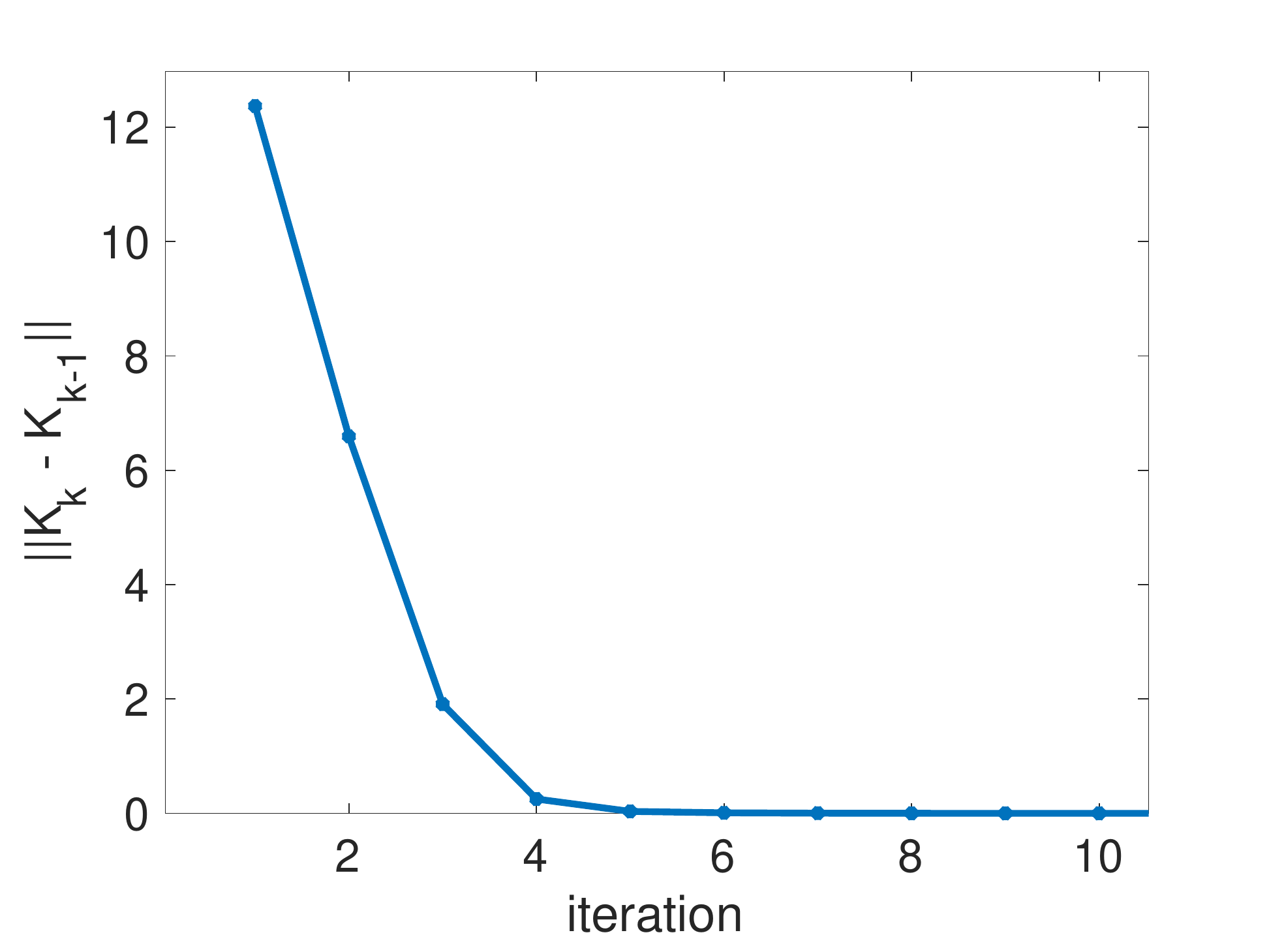} 
        \caption{Scenario A: K convergence}
        \label{fig:c1k}
    \end{minipage}
\vspace{-.4 cm}
\end{figure}

\footnotesize
\begin{align}
    K_{\mbox{unstruc}} = \begin{bmatrix}
    2.9234 &   0.7255  &  1.1487  &  0.3057  &  0.1397  &  0.2342 \\
    0.7255 &   2.6395  &  0.2282  &  0.0418  &  0.7435  &  1.0987\\
    1.1487  &  0.2282  &  2.9751 &   1.0436  &  0.0269  &  0.0547\\
    0.3057  &  0.0418  &  1.0436 &   4.0820 &   0.0001 &   0.0041\\
    0.1397  &  0.7435  &  0.0269 &   0.0001  &  3.2790 &   1.2881\\
    0.2342  &  1.0987  &  0.0547 &   0.0041  &  1.2881 &   2.7975
    \end{bmatrix},
\end{align}
\normalsize
with the objective of $12.0428$ units. 

\begin{figure}[!t]
    \centering
    \begin{minipage}{\linewidth}
        \centering
        \includegraphics[width = .75\linewidth, height =4.5 cm]{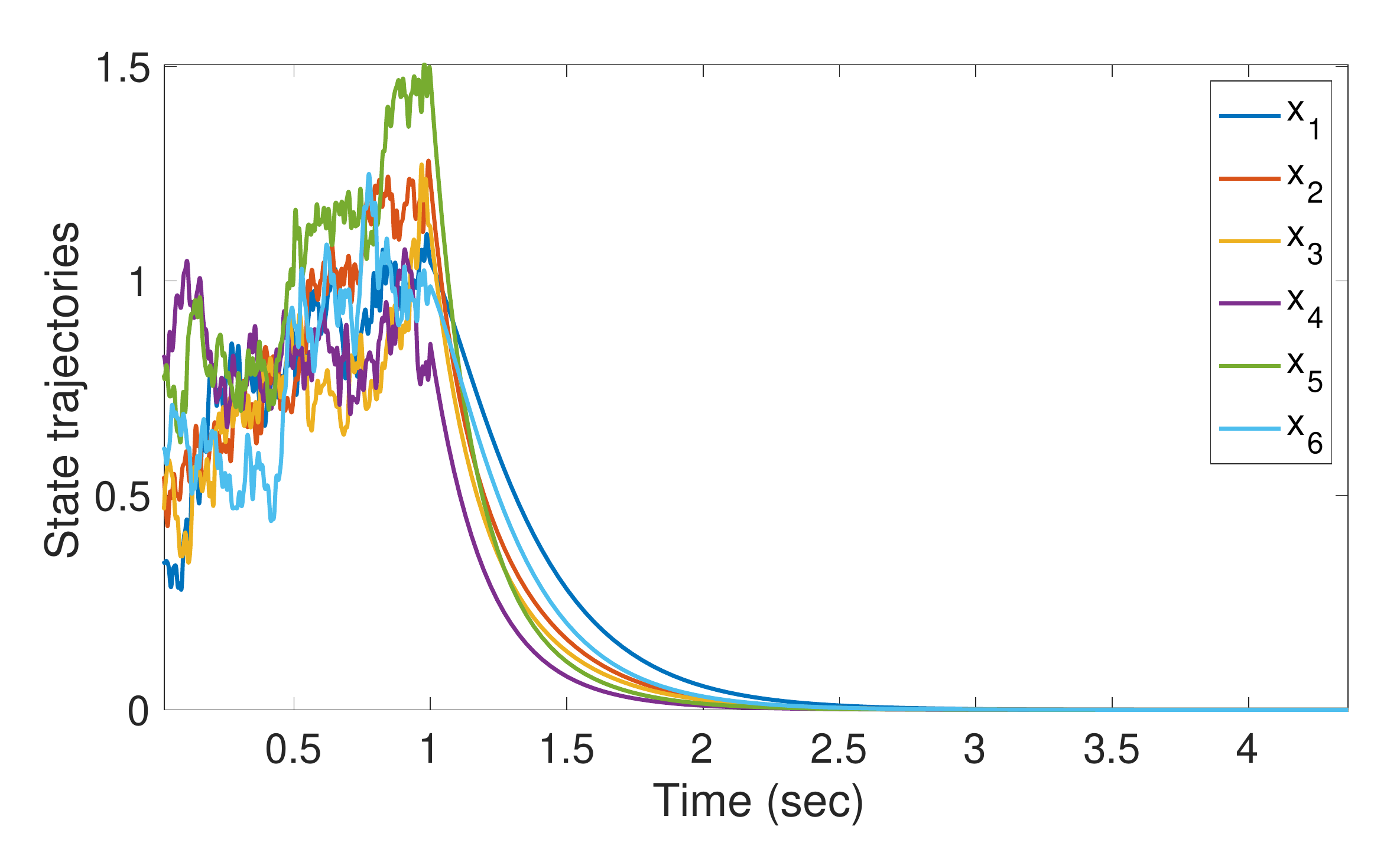} 
        \caption{Scenario B: State trajectories during exploration and control implementation}
        \label{fig:c2_traj}
    \end{minipage}
    \centering
    \begin{minipage}{0.4\linewidth}
        \centering
        \includegraphics[width = \linewidth]{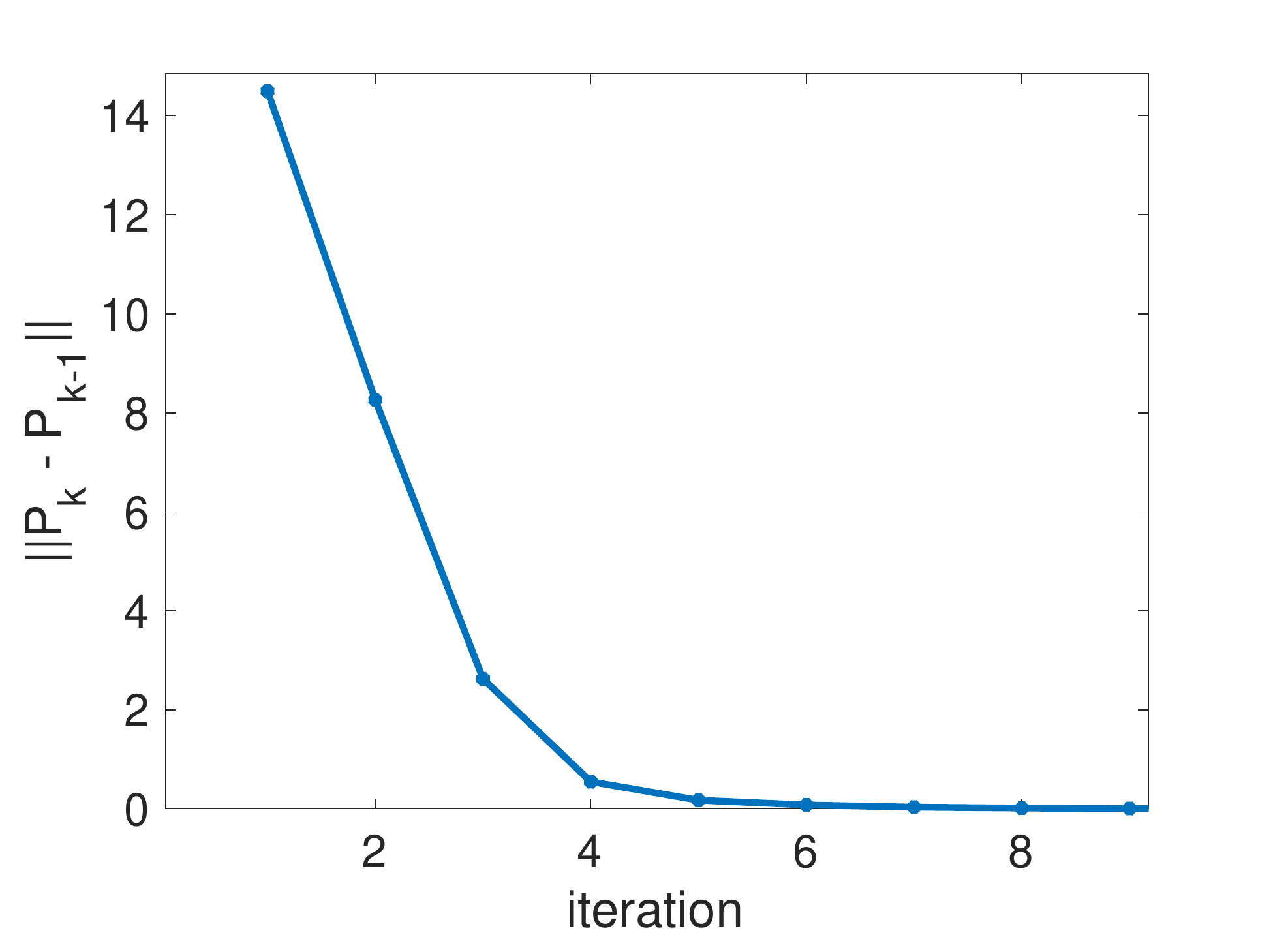} 
        \caption{Scenario B: P convergence}
        \label{fig:c2p}
    \end{minipage}
    \quad
    \begin{minipage}{0.4\linewidth}
        \centering
        \includegraphics[width = \linewidth]{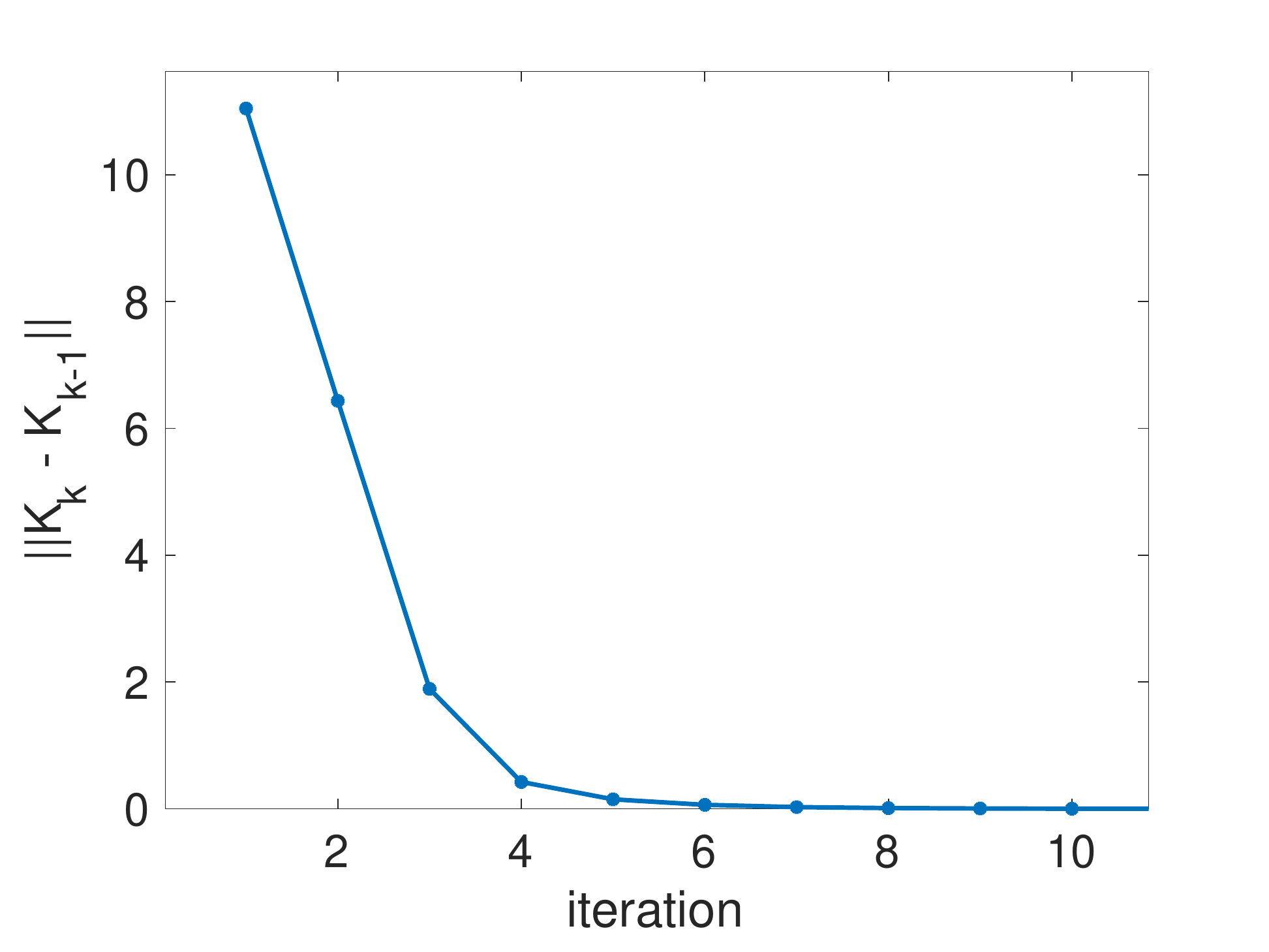} 
        \caption{Scenario B: K convergence}
        \label{fig:c2k}
    \end{minipage}
 \vspace{-.6 cm}
\end{figure}

We then perform similar experimentation with the scenario B, where we have removed more set of communication links. Here, the number of non-zero elements of $K$ is $23$. We need to gather at-least $88$ data samples, therefore, we perform around $1$ s of exploration with $0.01$ s time step. The structured control learned for this scenario is given by,

\footnotesize
\begin{align}
    K^B_{\mathcal{K}} = \begin{bmatrix} 
    0.0000 &   0.0000  &  1.2544  &  0.2898  &  0.1561&    0.0000\\
    1.0617  &  2.7750   & 0.2702  & 0.0000 &   0.7683  &  0.0000\\
    1.2544 &   0.2702 &   2.9979  &  0.0000 &   0.0000  &  0.0725\\
    0.0000 &  0.0000  &  1.0470 &   4.1729  &  0.0022 &   0.0046\\
    0.1561 &   0.7683 &  0.0000 &  0.0000 &   3.3002 &   1.3786\\
   0.0000  &  1.2338  &  0.0725  &  0.0000  &  1.3786 &  0.0000
    \end{bmatrix},
\end{align}
\normalsize
with the objective of $12.9764$ units. The state trajectories for scenario B during the exploration and the control implementation is given in Fig. \ref{fig:c2_traj}. The convergence of the least squre iterates for $P$ and $K$ are shown in Figs. \ref{fig:c2p}-\ref{fig:c2k}, where we can see that convergence is being reached after $8$ iterations. Also, the damping of the eigenvalues are improved with the control with the closed-loop eigenvalues are placed at $-10.61,
   -3.58,
   -4.22,
   -5.70,
   -9.19,
   -7.91$.

\vspace{-.5 cm}
\section{Conclusions and path forward}
This paper discussed model-free learning based computation of stable and close-to-optimal feedback gains which are constrained to a specific yet general structure. We first formulated a model-based condition to compute structured feedback control, which resulted in a modified algebraic Riccati equation (ARE). Techniques from dynamic programming and Lyapunov based stability criterion ensured the stability and the convergence. Thereafter, we embed the model-based condition into the a RL algorithm where trajectories of states and controls are used to compute the gains. The algorithm uses an interleaved policy evaluation and policy improvement steps. We also analyzed the sub-optimality of the structured optimal control solution in comparison with the unstructured optimal solutions. Simulations on a multi-agent network with constrained communication infrastructure validated the theoretical results. 
Our future work will  investigate the possibility to embed a prescribed degree of stability margin along with the structural constraint, as well as necessary and sufficient conditions for the existence of the optimal structured control.    

\bibliographystyle{IEEEtran}
\bibliography{ref}
\end{document}